\documentclass[useAMS,usenatbib]{mn2e}
\usepackage{times,pslatex}
\usepackage{amsmath,amsfonts,graphicx}

\def\hr{HR~8799}
\def\idm#1{{\mbox{\scriptsize #1}}}
\def\Ym{\left<Y\right>}

\newcommand\Chi{{(\chi^2_\nu)^{1/2}}}

\def\idm#1{{\mbox{\scriptsize #1}}}

\newcommand{\mJ}{\mbox{m}_{\idm{Jup}}}
\newcommand{\MS}{\mbox{M}_{\odot}}
\title[Planetary scattering in the HR~8799 system?]
{Is the HR~8799 extrasolar system destined for planetary scattering?}
\author[K. Go\'zdziewski \& C. Migaszewski ]{
Krzysztof Go\'zdziewski\thanks{E-mail:k.gozdziewski@astri.umk.pl} \&
Cezary Migaszewski\thanks{E-mail: c.migaszewski@astri.umk.pl}\\
Nicolaus Copernicus University, Gagarin Str. 11, 87-100 Toru\'n, Poland
}
\begin{document}
%
\date{Accepted 2009 April 16.  Received 2009 April 14; in original form 2009
March 13}
\pubyear{2008}
\maketitle
\label{firstpage}
%
\begin{abstract}
The recent discovery of a three-planet extrasolar system of \hr{} by Marois et
al. is a breakthrough in the field of the direct imaging. This great achievement
raises questions on the formation and dynamical stability of the \hr{} system, because
Keplerian fits to astrometric data are strongly  unstable during $\sim 0.2$~Myr. We search for
stable, self-consistent $N$-body orbits with the so called GAMP method that
incorporates stability constraints into the optimization algorithm.  Our
searches reveal only small regions of stable motions in the phase space of
three-planet, coplanar configurations. Most likely, if the planetary masses are
in 10-Jupiter-mass range, they may be stable only if the planets are involved in
two- or three-body mean motion resonances (MMRs). We found that 80\% systems
found by GAMP that survived 30~Myr backwards integrations, eventually become
unstable after 100~Myr. It could mean that the HR~8799 system undergo a phase of
planet-planet scattering. We test a hypothesis that the less certain detection
of the innermost object is due to a blending effect. In such a case, two-planet
best-fit systems are mostly stable, on quasi-circular orbits and close to the
5:2~MMR,  resembling the Jupiter-Saturn pair.
\end{abstract}
%
\begin{keywords}
stars: individual: HR~8799; methods: numerical; methods: N-body simulations 
\end{keywords}

\section{Introduction}
The \hr{} planetary system was detected by \cite{Marois2008} through the direct
imaging. Soon, a new observation was added by \cite{Lafreniere2009} who
reanalyzed images done in 1998, extending the observational window to $\sim
10$~years and four different epochs. [We skip the most recent observation in
\citep{Fukagawa2009} that appeared after we finished this paper, because it did
not change the initial condition]. Although the semi-major axes are large (about
of 24, 36 and 68~au, respectively), the massive companions strongly interact
mutually. As we show, their orbits remain in extremely chaotic zone spanned by
low-order MMRs.  We attempt to constrain the initial conditions by available
astrometric data and seemingly obvious requirement of astronomical stability.
Our work  complements papers of \cite{Fabrycky2008} and \cite{Reidemeister2009}.
Here,  we follow a different approach that relies on quasi-global,
self-consistent search for stable best-fit systems, the so called GAMP
\citep[e.g.,][]{Gozdziewski2008} which was used to model the radial velocity
data. The direct imaging seems also a particularly good target for this 
numerical technique. 

\begin{figure*}
\centerline{
 \vbox{
        \hbox{\includegraphics[width=54mm]{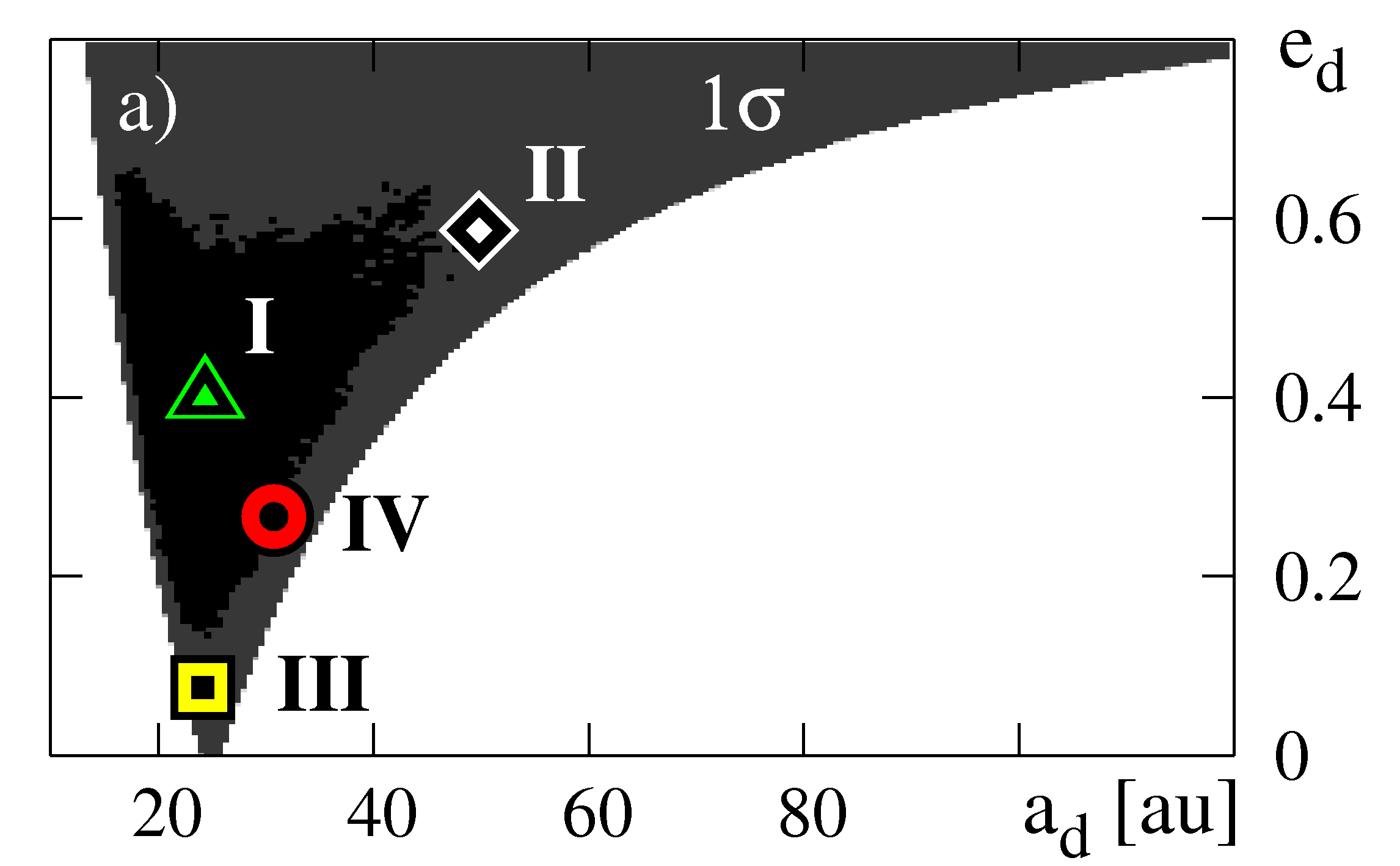} 
              \includegraphics[width=54mm]{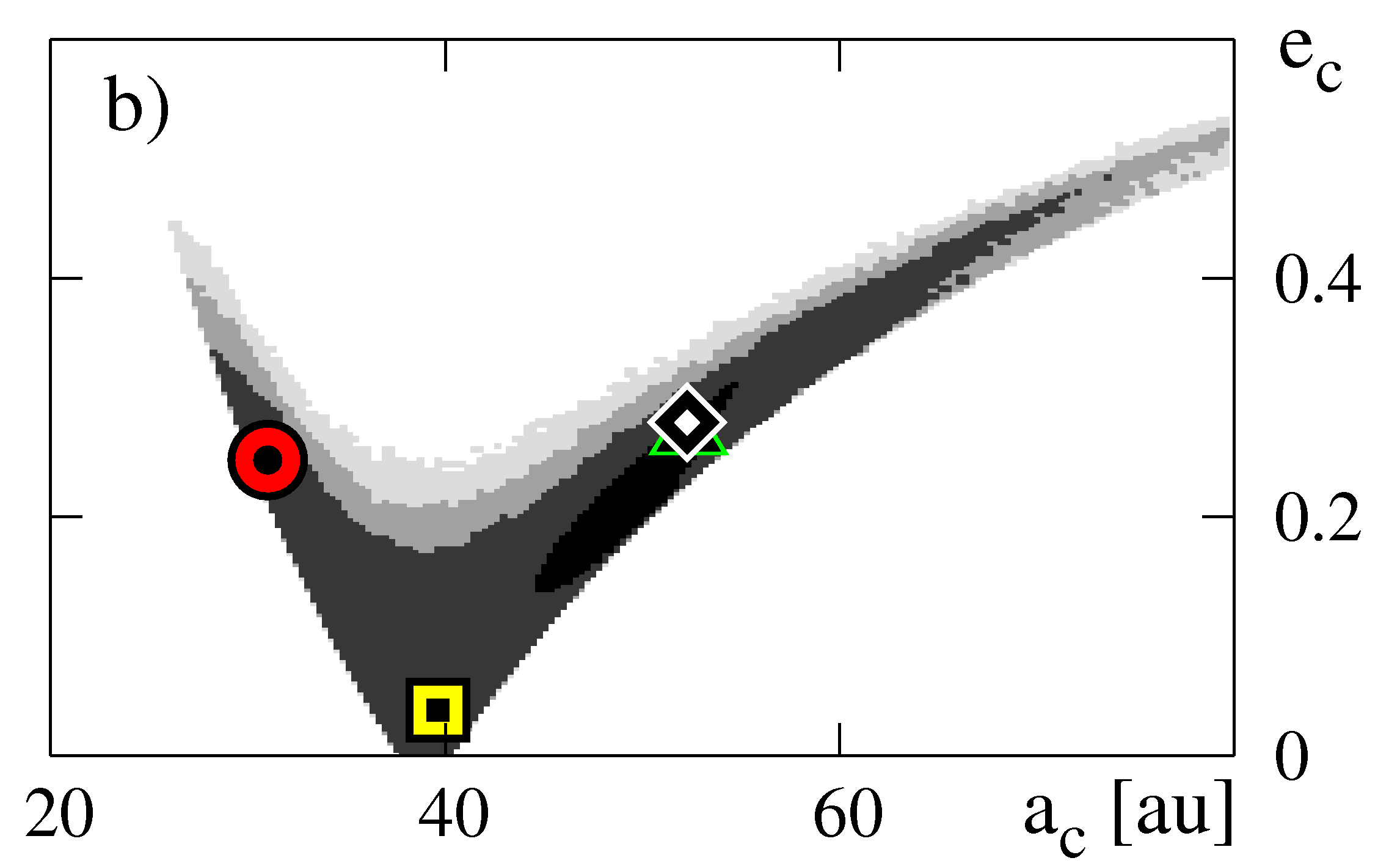} 
	      \includegraphics[width=54mm]{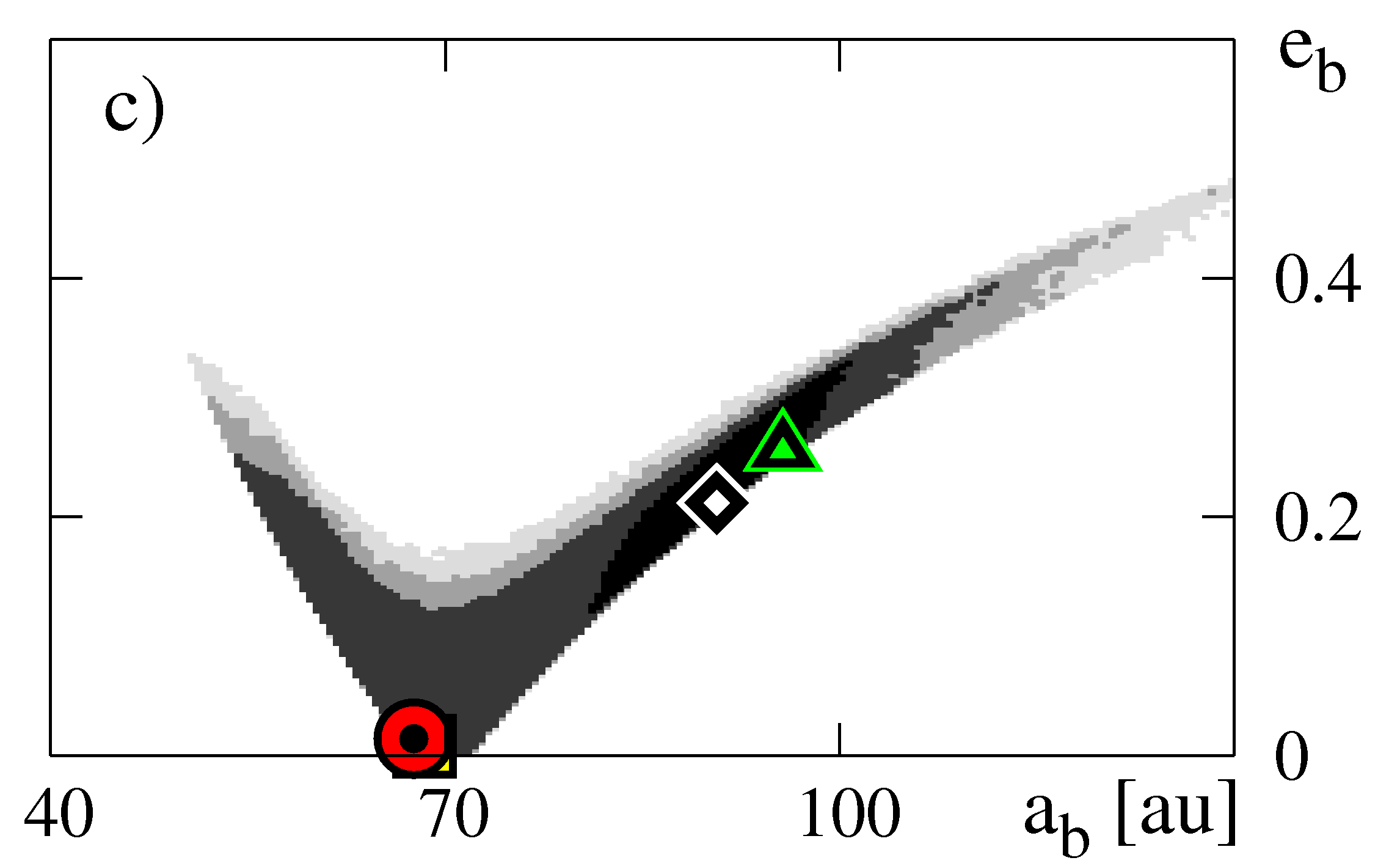}
              }
        \hbox{\includegraphics[width=54mm]{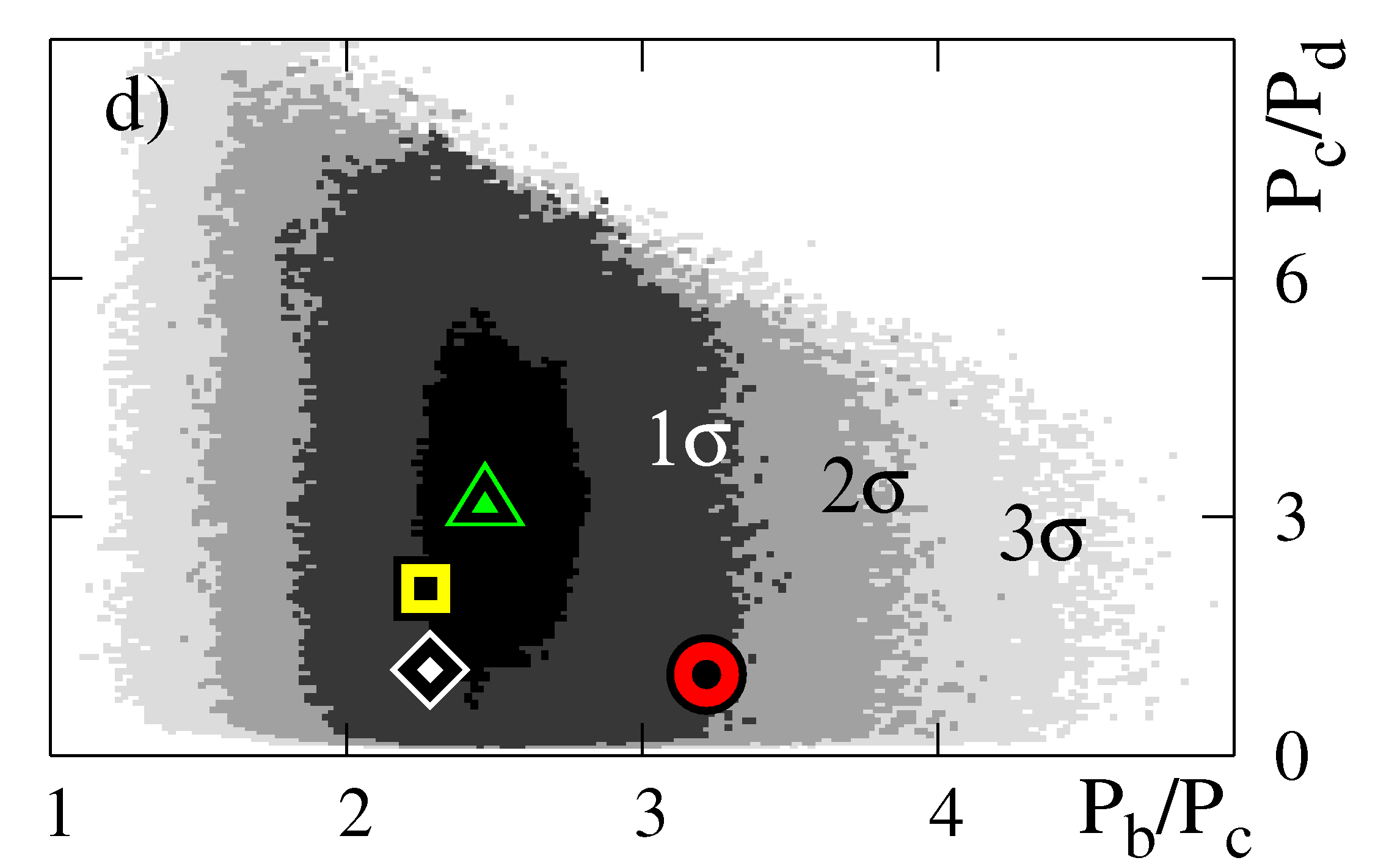}
              \includegraphics[width=54mm]{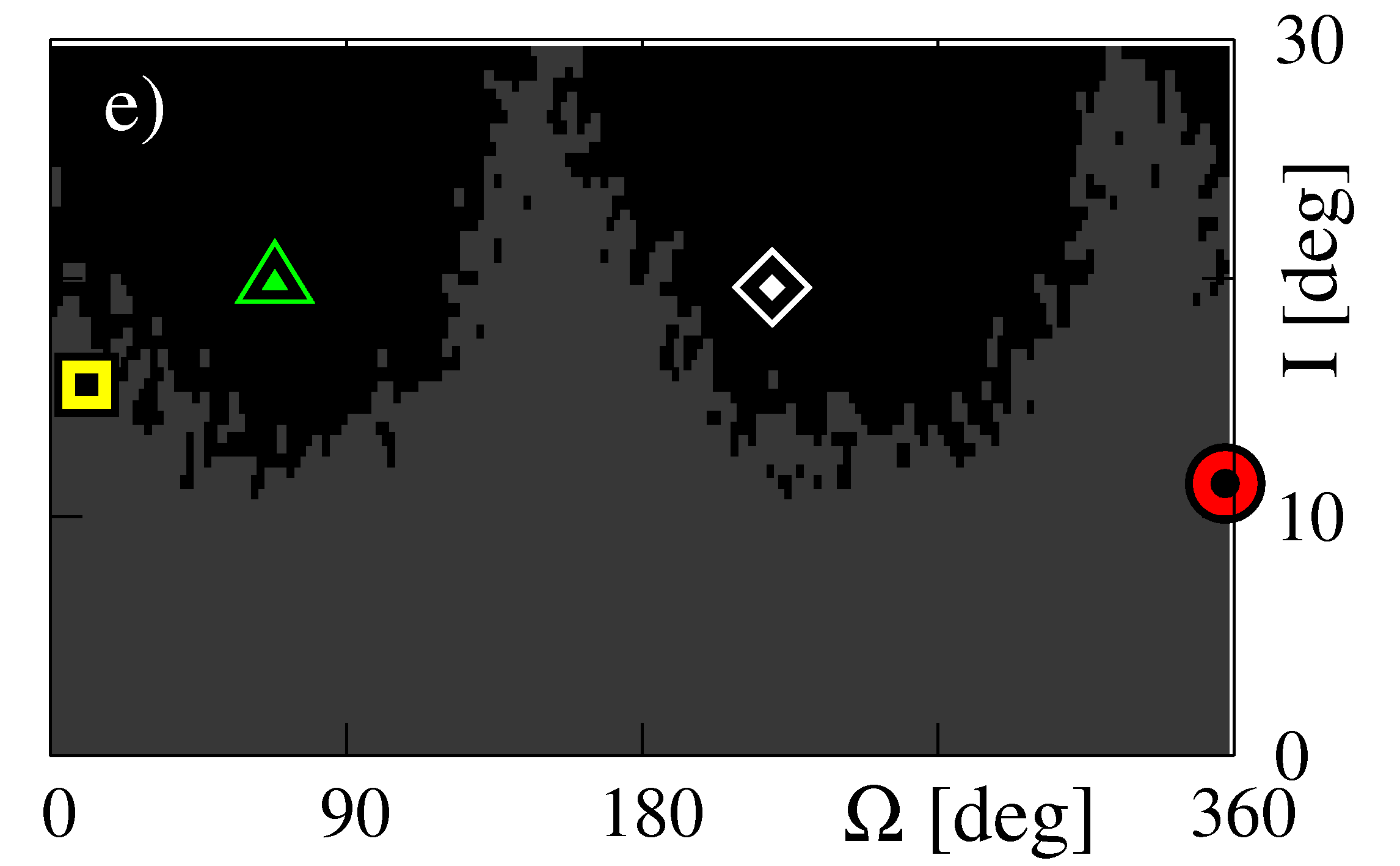}\
	      \includegraphics[width=54mm]{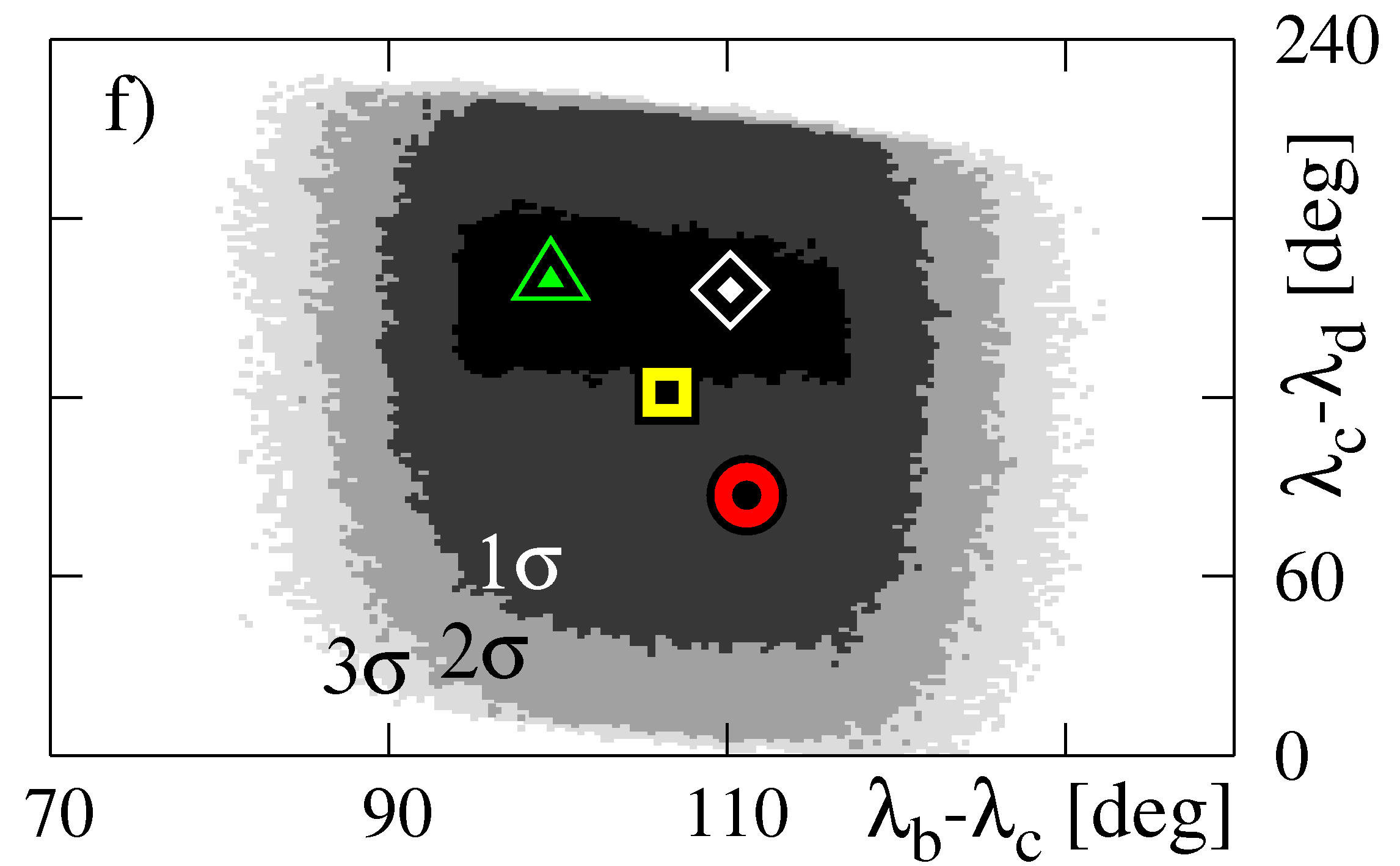}
              }
  }
}
\caption{
Global topology of $\Chi$ projected onto selected planes of Keplerian, 
osculating elements at the epoch of Sept.~18, 2008.  Symbols are for the
best-fits systems analyzed in this work.  They are labeled accordingly with
Table~1. Shaded areas are for $1\sigma$, $2\sigma$, and $3\sigma$-levels labeled
in panels~(a,d,f) [$\Chi<1.67$, $\Chi<1.85$, and $\Chi<2.09$, respectively]  of
the best Fit~I marked with green triangle; darker shade means better fit. Black
regions are for $\Chi$ only marginally worse from the best-fit value. Other fits
analyzed in this work are marked with white diamond (Fit II, unstable Trojans),
yellow rectangle (Fit~III, the best-fit stable configuration), and red circle
(Fit~IV, stable Trojans). See  Fig.~\ref{fig:fig2} for a geometry of these
solutions.
}
\label{fig:fig1}
\end{figure*}
Following astrometric estimates of the semi-major axes, we see that the 
observational window covers a tiny part of orbital periods which are counted in
hundreds of years. The initial condition may be determined with a significant
error. To illustrate this uncertainty, we map the multi-cube of astrometric
coordinates $x(t),y(t)$ and velocities $v_x(t),v_y(t)$ [from the slope of
$x(t),y(t)$] within $3\sigma$-level of the linear model of $x(t),y(t)$ onto
osculating Keplerian elements at the epoch of Sept.~18, 2008. This  most
reasonable choice follows the very short time-span of observations. The data set
consists of 13 mean positions in $[E,N]$-axes in \citep{Marois2008} as well as one 
observation in \citep{Lafreniere2009}; we also adopted a standard HIPPARCOS
distance to the star of $39.4 \pm 1.1$~pc.

The astrometric model is parameterized by the stellar mass $m_0$, $N$ tuples of
Keplerian elements ${\bf p}_p$ = ($m$[m$_{\idm{Jup}}$], $a$[au], e, $\omega$[deg],
{${\cal M}_0$[deg]), i.e., the mass, semi-major axis, eccentricity, argument of
pericenter, and the mean anomaly (or longitude $\lambda$), for each planet $p=1,\ldots,N$, respectively,
and two Euler angles describing the inclination ($i$) and the nodal longitude
($\Omega$) of the orbital plane with respect to the plane of the sky. The
$\Chi$-function is build up from deviations of  astrometric measurements from
coplanar, projected orbits. It depends indirectly on the astrophysical mass
constraints through the transformation of the velocity--Keplerian elements.
Following \cite{Marois2008}, the planetary masses are: 
m$_{\idm{b}}=7_{-2}^{+4}$~m$_{\idm{Jup}}$, m$_{\idm{c}}=10 \pm 3$~m$_{\idm{Jup}}$,
m$_{\idm{d}}=10 \pm 3$~m$_{\idm{Jup}}$; the mass of the parent star is $m_0=1.5
\pm 0.3$~m$_{\sun}$. They are roughly consistent with the  recent, independent
estimates of \cite{Reidemeister2009}. In our simulations, all {\em masses are
free parameters} which are varied within their $1\sigma$-error ranges.
(Moreover, a proper mass determination may be critically important for the
stability analysis). Figure~\ref{fig:fig1} shows levels of $\Chi$ in selected
two-dimensional planes of osculating elements at the epoch of Sept.~18, 2008.
The best-fit solution is marked with a green triangle. 
{
It is roughly consistent with a model of not too eccentric, face-on orbits by
the discovery team.
We found that the} limited astrometric data permit a continuum of models with
different orbital characteristics, e.g., eccentricities within $1\sigma$-level
of the best fit may be as large as $\sim 0.4$. The orbital periods consistent
with relatively small $\Chi$ may be found in a proximity of numerous low-order  MMRs. In
turn, these factors strongly affect the dynamical stability of the system. 

The geometry of the nominal, best-fit solution with $\Chi \sim 1.55$,
over-plotted on the original image,  is illustrated in 
Fig.~\ref{fig:fig2}(I). This best fit system appears strongly unstable and
self-disrupts after $\sim 0.2$~Myr, so our conclusion is the same as in
\citep{Fabrycky2008}. Moreover, according with our Fig.~\ref{fig:fig1}d, very
different orbital solutions are possible. For instance, two inner planets might
be involved in 1c:1d~MMR, or other low-order MMRs.  An example of unusual Trojan
configuration with only marginally worse $\Chi \sim 1.56$ from the best-fit
model is shown in Fig.~\ref{fig:fig2}(II). Also such  ``raw'', kinematic fits
are catastrophically unstable during the first Myr. Curiously, in these
solutions, $m_0$ tends to the lowest possible limit that might indicate an
internal inconsistency of the model with the data, if we recall that  the
stellar mass is constrained {\em a priori}.

\begin{figure*}
\centerline{
        \hbox{\includegraphics[  width=43mm]{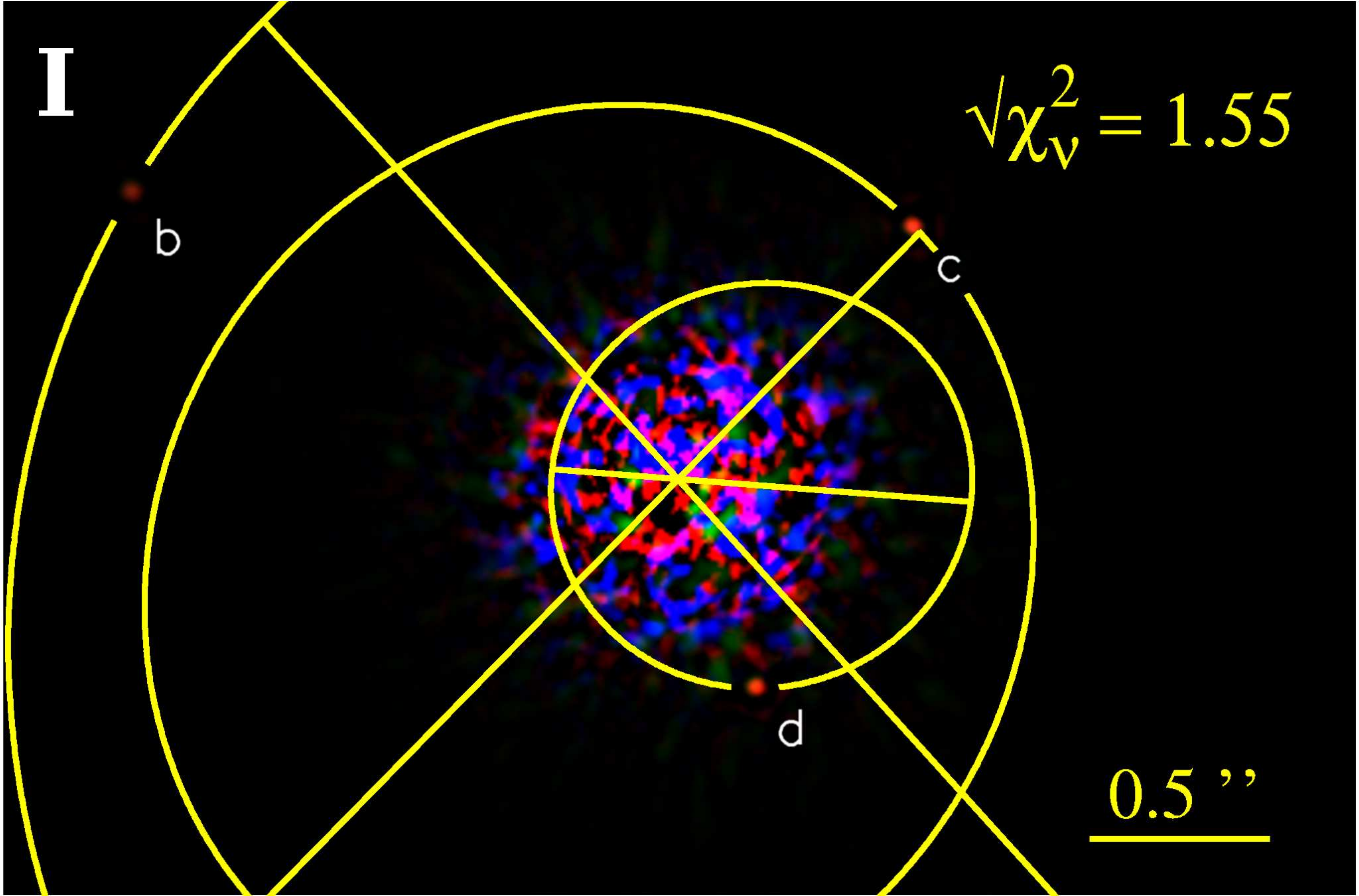}\hspace*{0mm} 
              \includegraphics[  width=43mm]{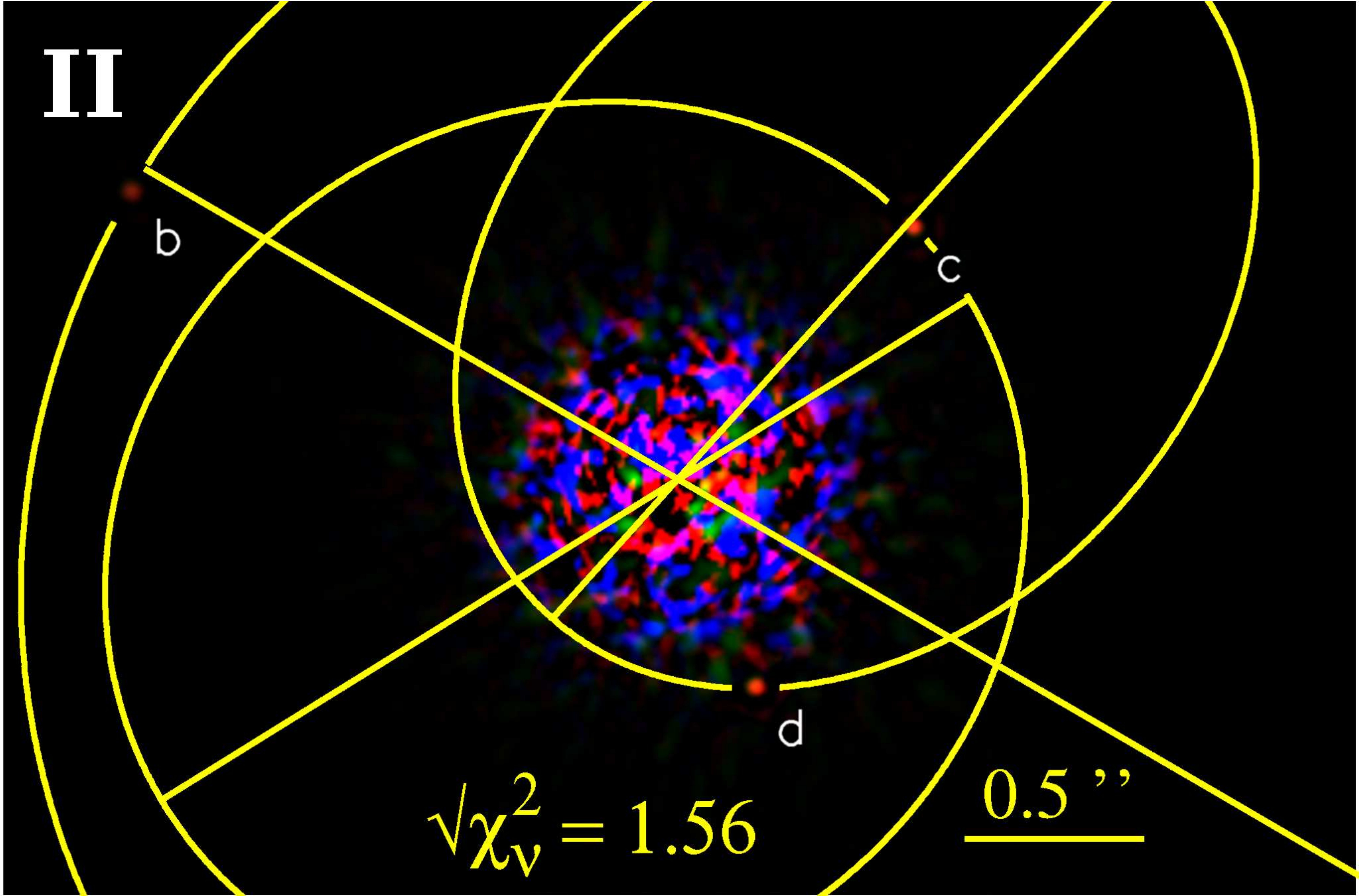}\hspace*{0mm}
	      \includegraphics[  width=43mm]{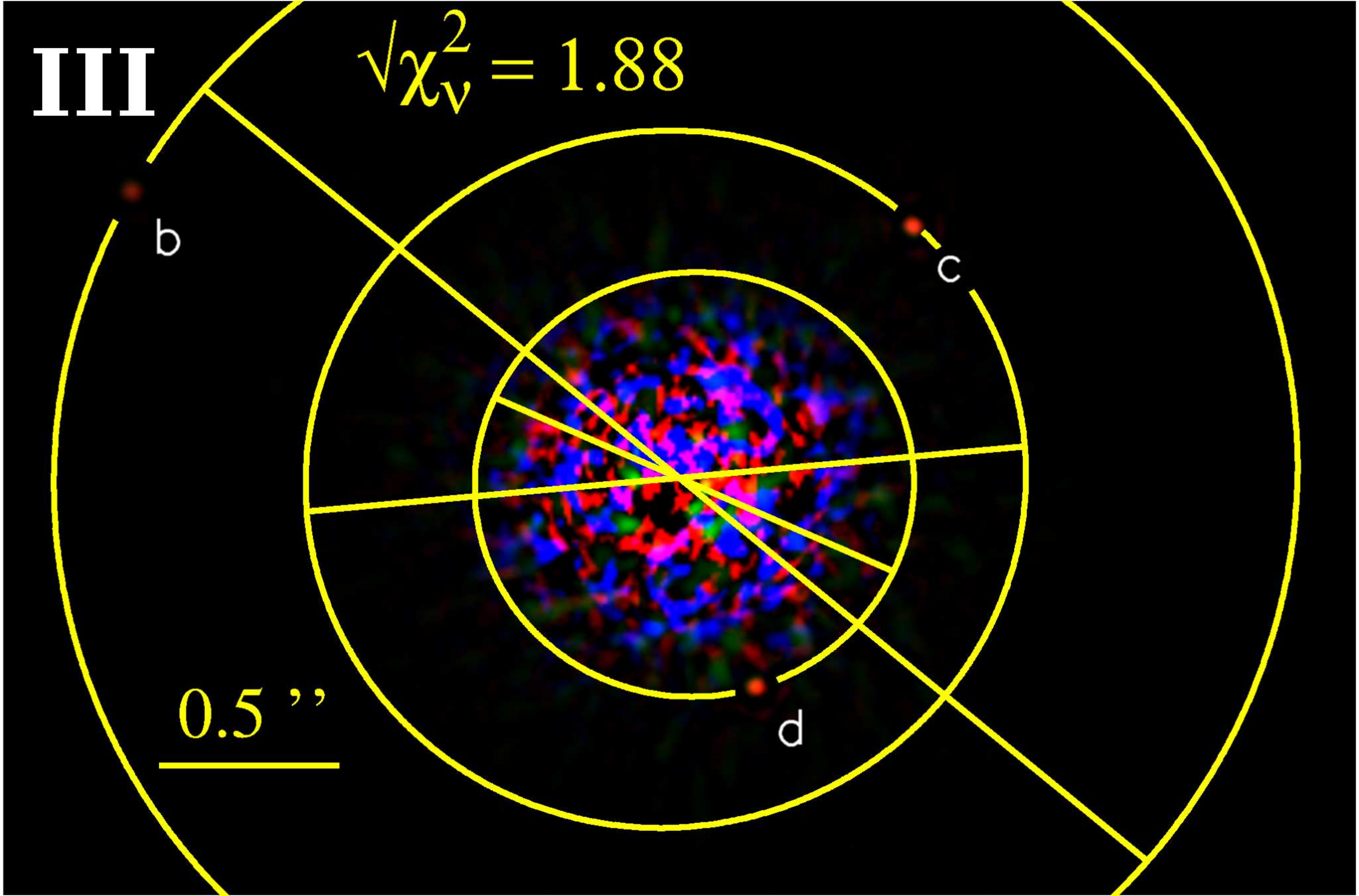}\hspace*{0mm}
	      \includegraphics[  width=43mm]{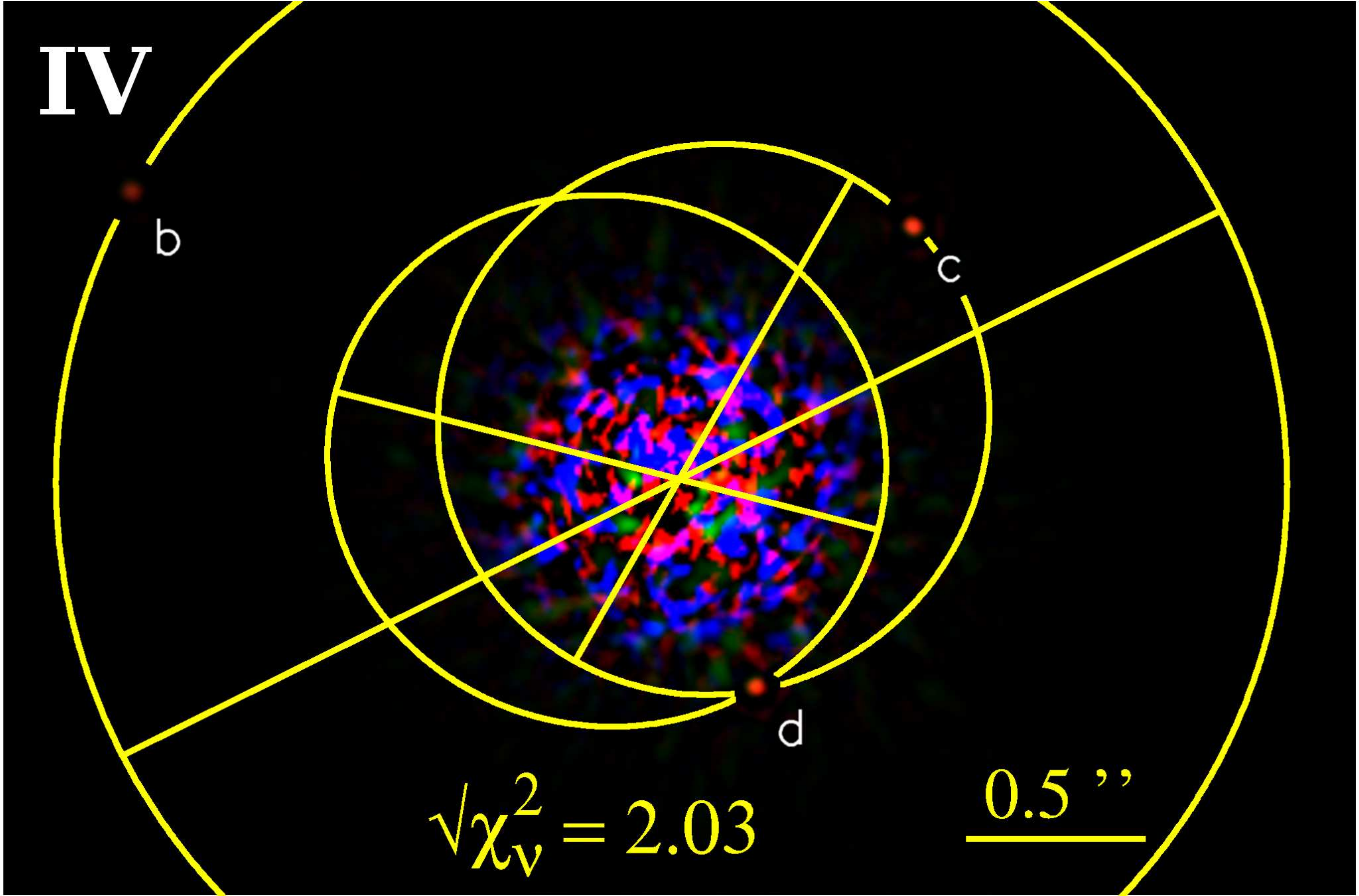}}
}
\caption{
Orbital geometry of the best-fit configurations  projected onto the plane of the
sky and the true image of the system by combined photographs taken with the
Keck~II adaptive optics.  The planets appear as red dots around the residual
scattered light of the star (seen in the center). The best-fit osculating orbits
at the epoch of Sept.~18, 2008 are drawn with yellow ellipses. Stright lines are
for  the apsidal lines of these orbits. The HR~8799 image credit:
NRC Canada/C. Marois. See Table~1 for the  osculating
elements of these fits labeled accordingly.
}
\label{fig:fig2}
\end{figure*}

\begin{table*}
\begin{tabular}{c c | c c c c c | c c c c}
\hline
Fit & Planet & $m~[\mJ]$ & $a$~[au] & $e$ & $\omega$~[deg] & $\mathcal{M}_0$~[deg] & $m_0~[\MS]$ & $i$~[deg] & $\Omega$~[deg] & $\Chi$ \\
\hline
I  & b & $6.509$ & $95.680$ & $0.255$ & $65.242$ & $11.04$ & & & & \\
kinematic & c & $12.57$ & $52.364$ & $0.269$ & $335.95$ & $0.740$ & $1.200$ &
$19.8$ & $68.3$ & $1.55$ \\
unstable  & d & $12.10$ & $24.420$ & $0.399$ & $106.32$ & $71.08$ & & & & \\
\hline
II   & b & $6.108$ & $90.676$ & $0.212$ & $289.28$ & $1.480$ & & & & \\
kinematic & c & $9.644$ & $52.275$ & $0.279$ & $171.84$ & $8.686$ 
& $1.225$ & $19.6$ & $219.5$ & $1.56$ \\
unstable   & d & $7.806$ & $49.855$ & $0.587$ & $9.0134$ & $15.42$ & & & & \\
\hline
III    & b & $8.022$ & $68.448$ & $0.008$ & $308.59$ & $191.7$ & & & & \\
GAMP   &  c & $11.87$ & $39.646$ & $0.012$ & $353.83$ & $40.03$ & $1.445$ & 
$15.5$ & $11.12$ & $1.88$ \\
 stable 1d:2c:4b~MMR  & d & $8.891$ & $24.181$ & $0.075$ & $144.38$ & $127.6$ & & & & \\
\hline
 IV  & b & $9.708$ & $67.661$ & $0.014$ & $29.671$ & $123.8$ & & & & \\
GAMP & c & $7.963$ & $31.045$ & $0.248$ & $243.41$ & $158.8$ & 
$1.611$ & $11.4$ & $357.2$ & $2.03$ \\
stable 1d:1c~MMR  & d & $7.397$ & $30.777$ & $0.267$ & $348.38$ & $326.5$ & & & & \\
\hline
V& b & $8.325$ & $73.543$ & $0.043$ & $122.49$ & $357.8$ & 
$1.448$ & $17.9$ & $30.9$ & $1.48$ \\
GAMP (stable $\sim$ 2c:5b~MMR)  & c & $9.011$ & $39.358$ & $0.076$ & $307.34$ & $61.35$ & & & & \\
\hline
\end{tabular}
\caption{
Osculating elements of the best-fit solutions at the epoch of Sept.~18, 2008.
Formal errors of kinematic Fit~I and~II can be estimated  graphically, see
Fig.~\ref{fig:fig1}. Note that formal errors of dynamical GAMP Fits~III and IV
can be taken  the same as for Fits~I and~II but the error ranges are strictly
limited to stable regions of the phase space. The same concerns Fit~V (see the
text).  The orbital inclination and nodal longitude  (Fig.~\ref{fig:fig1}e) and
arguments of pericenter (not shown here) are unconstrained. We set $i\in
[0,30^{\circ}]$, consistent with estimates of \citep{Marois2008}, following the
rotation model of the star. 
}
\end{table*}

\vspace*{-4mm}
\section{The best-fit stable configurations}

Recent works \cite[e.g.,][and references
therein]{Juric2008,Chatterjee2008,Scharf2009} showed that compact planetary
systems may evolve towards configurations spanning  wide ranges  of orbital
elements. The long-period planets could place constraints on early stage planet
formation scenarios. An interpretation of the direct imaging surveys is also closely
related to models of the dynamical relaxation \citep{Veras2009}. Hence, even
apparently odd solutions (like the Trojan configurations), consistent with
observations,  should not be skipped {\em a priori}. Because the parent star may
be very young (30~Myr or less), and the dynamical separation of planets in terms
of the mutual Hill radii, \citep{Chatterjee2008}, $K \sim 2$, the three-planet
system is {\em strongly unstable} in a few hundred orbital periods time-scale
(see their Fig.~29; although these calculations are for more compact,
Solar-system like models and planets in Jupiter mass range). So the \hr{} system
might be not yet dynamically relaxed, remaining in a stage of planet-planet
scattering. On the other hand, we may be ``fooled'' by the significant errors of
the initial condition implied by short time-base of the astrometry. Then the
requirement of the dynamical stability may help us to find long-living systems
close to apparently unstable best-fit configurations.   

As is well known, the phase-space of a compact multi-planet system has 
non-continuous structure with respect to any notion of stability.   The
permitted region in the 18-dimensional parameter space of the \hr{} system is
large and has complex shape. To explore it efficiently, we apply a variation of the GAMP method
\citep[see e.g.,][for details]{Gozdziewski2008} which relies on self-adapting
optimization based on the genetic algorithms (GAs)
\cite[e.g.,][]{Charbonneau1995,Deb2002} and on ``penalizing'' unstable
configurations by an appropriate term added to the mathematical value of $\Chi$.
Here, the penalty term is expressed through the diffusion of fundamental
frequencies \citep{Robutel2001,Nesvorny1996}. 

An extensive GAMP search revealed long-term stable best-fit~III (Table~1)
illustrated in Fig.~\ref{fig:fig2}(III). We note that its $\Chi \sim 1.88$,
remaining within $2\sigma$-range of the nominal, kinematic Fit~I. To understand 
this solution, we computed its dynamical maps in terms of the Spectral Number
(SN), the fast indicator invented by \cite{Michtchenko2001},  and the $\max e$
(the maximal eccentricity attained during prescribed integration time). The SN
map is illustrated in Fig.\ref{fig:fig3}. Fit~III lies inside a small island of
regular motions (its width for the innermost planet is only $\sim 0.3$~au). A
map of the $\max e$ indicator  (not shown here) reveals that outside this
region, one of orbits become highly eccentric that leads to catastrophic events
during $\sim 1$~Myr.  Fit~III  describes a configuration involved in the
Laplace-type three-body resonance, 1d:2c:4b~MMR. Its critical argument is shown
in the middle panel of Fig.~\ref{fig:fig4}. A similar solution was already found
by \cite{Fabrycky2008} and analyzed in more detail by \cite{Reidemeister2009}. 
Actually, our Fit~III is also unstable but on a very long time-scale. After
$\sim 400$~Myr,  the innermost eccentricity suddenly grows and the Laplace
resonance disrupts (see two upper panels in Fig.~\ref{fig:fig4}), indicating a
collision. Hence, the small amplitude of the resonance angle does not protect
the system from the collision. In fact, Fit~III is formally chaotic that is
indicated by the MEGNO  in  Fig.~\ref{fig:fig4}. This shows that the stability
depends on long-term effects of the three-body mutual interactions and is
tightly related to formally chaotic or regular character of  tested
configurations.
\begin{figure}
\centerline{
        \hbox{\includegraphics[width=78mm]{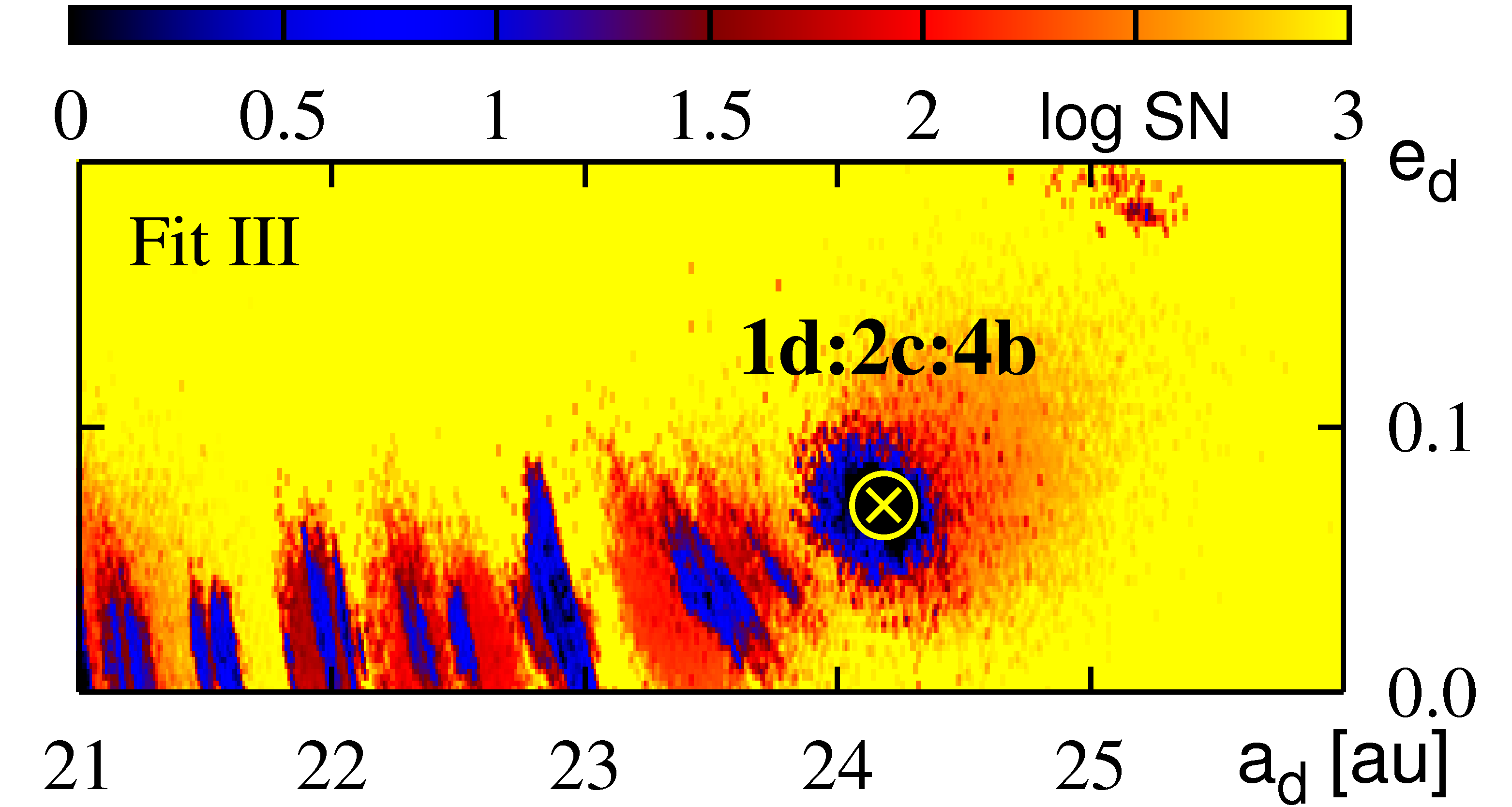}}
}
\caption{
The SN map around stable Fit~III (yellow rectangle in Fig.~\ref{fig:fig1}) in
the $(a_{\idm{c}},e_{\idm{c}})$-plane; other elements are fixed at their nominal
values (see Table~1). Yellow colour is for strongly chaotic systems, black is for stable
solutions. The maximal integration time for each pixel is $10$~Myr
($\sim 25000 P_{\idm{d}}$).
}
\label{fig:fig3}
\end{figure}

Besides Fit~III, we also found a {\em stable} fit related to 1c:1d~MMR, with
moderate $e_{\idm{c,d}}\sim 0.25$ and $a_{\idm{c,d}}\sim 31$~au, see
Fig.~\ref{fig:fig1},\ref{fig:fig2}(IV). Its $\Chi \sim 2$  is still acceptably
small because it lies within the formal $3\sigma$-level  of the best Fit~I. The
dynamical SN map of this fit is shown in Fig.~\ref{fig:fig5}. It reveals also a
small island of stable motions having  the width comparable to Fit~III.
Simultaneously, planet~b remains in a narrow island close to (1c:1d):3b~MMR. The
Trojans live at least over 3~Gyr ---  this system is close to quasi-periodic
one, as indicated by $\Ym{}(t) \sim 2$ \citep{Cincotta2003} over large part of
the integration time (Fig.~\ref{fig:fig6}).  Still, a 10~Myr MEGNO map (not
shown here) shows that the island of regular solutions is very tiny ($\sim
0.01$~au). This fit is also weakly-chaotic although  during first
600~Myr it appears as regular. This solution has peculiar small-amplitude
librations of apsidal angle $\Delta\varpi(t)=\varpi_{\idm{d}}-\varpi_{\idm{c}}$
around $100^{\circ}$ (the upper panel in Fig.~\ref{fig:fig6}). It might be the
first case of {\em asymmetric} librations in the 1:1~MMR  observed in a real
system, and  predicted already in low-order resonances \citep[in particular, in
2:1~MMR,][]{Hadjidemetriou2006}.
\begin{figure}
\centerline{
        \hbox{
	      \includegraphics[width=70mm]{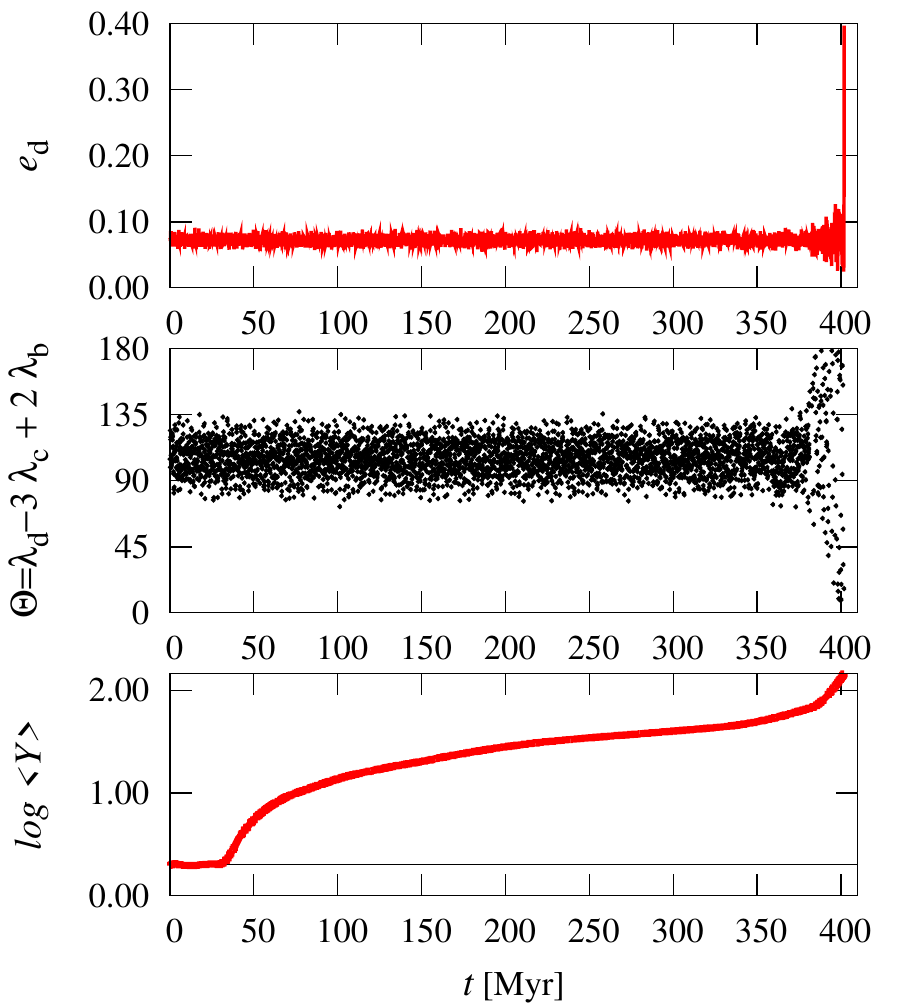}}
}
\caption{
The innermost eccentricity in Fit~III (the top panel), the argument of the
Laplace resonance $\theta=\lambda_{\idm{d}} - 3\lambda_{\idm{c}}  +
2\lambda_{\idm{b}}$ (the middle panel) and MEGNO, $\Ym(t)$ (the bottom panel).
$\Ym(t)$ converges to 2 for regular systems and diverges linearly for chaotic
motions \citep{Cincotta2003}.
}
\label{fig:fig4}
\end{figure}

\begin{figure}
\centerline{
        \hbox{\includegraphics[width=74mm]{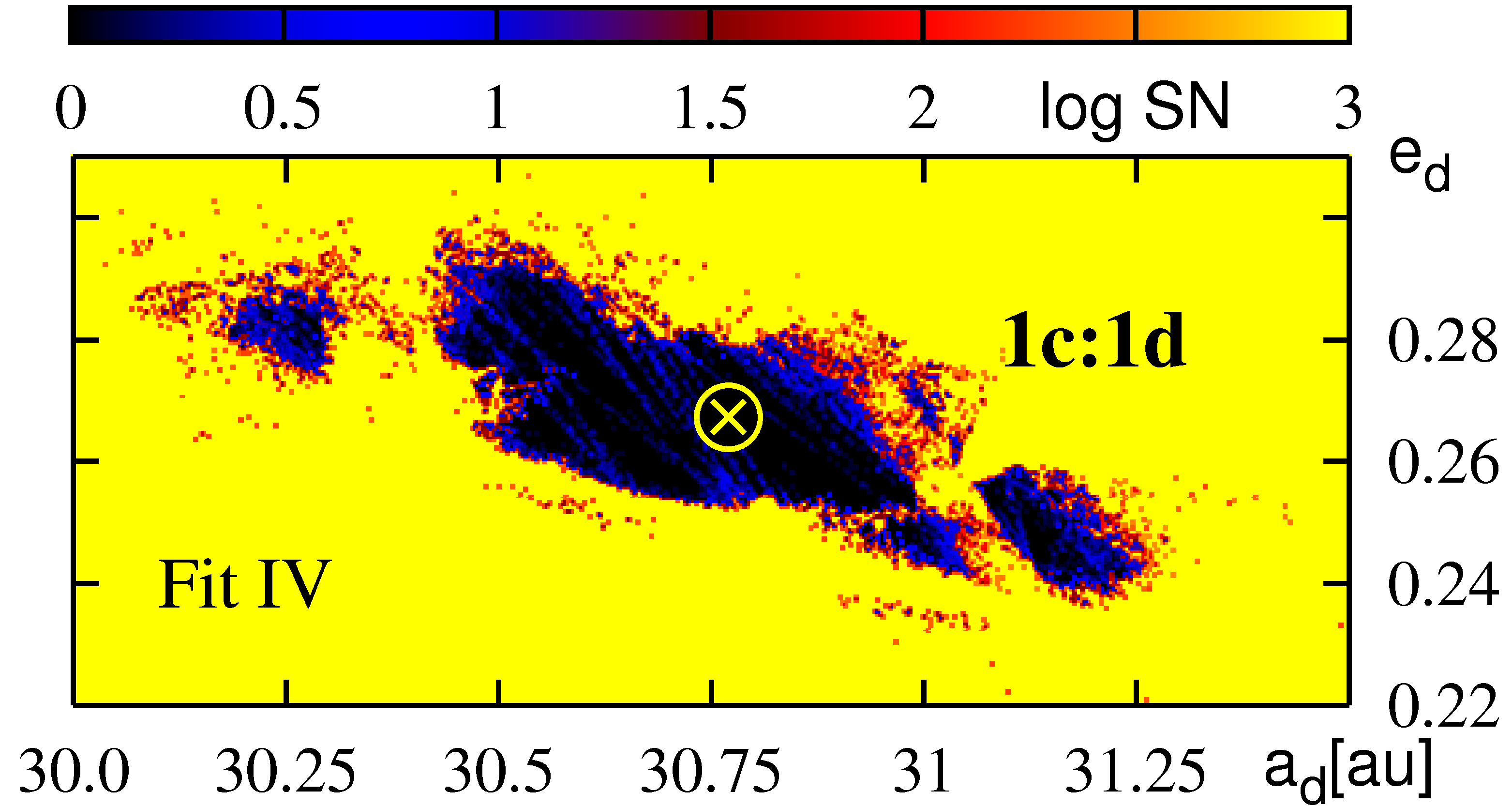}} 
}
\caption{Dynamical SN map of a stable 1c:1d~MMR (Fit~IV, see Table~1).
}
\label{fig:fig5}
\end{figure}

\begin{figure}
\centerline{
      \hbox{\includegraphics[width=70mm]{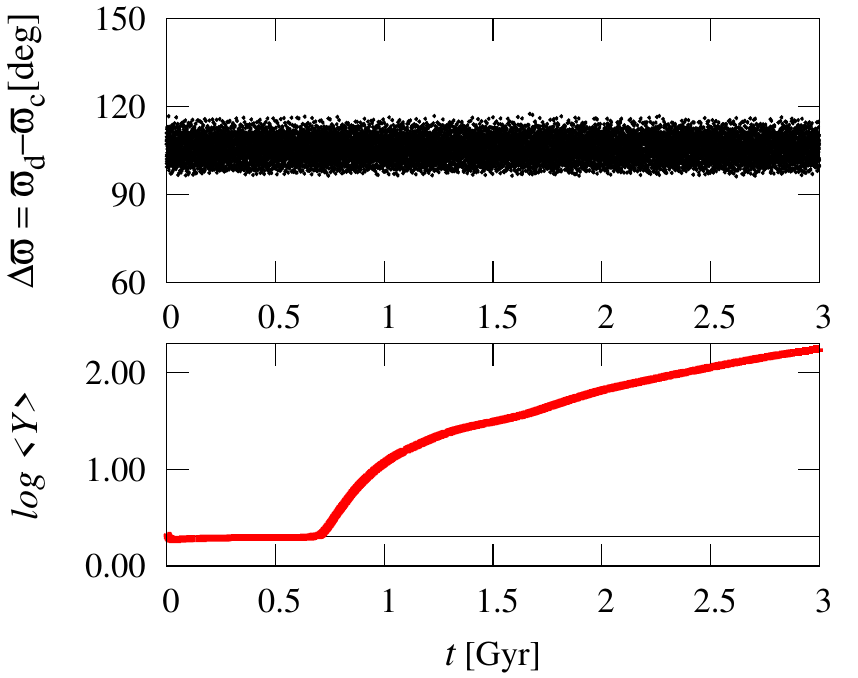}}
}
\caption{
Long-term stability of the best-fit solution with two inner planets involved in
1:1~MMR (Fit IV, see Tab.~1).  The upper panel is for apsidal angle 
$\Delta\varpi=\varpi_{\idm{d}}-\varpi_{\idm{c}}$. The bottom panel is for the
MEGNO indicator.
}
\label{fig:fig6}
\end{figure}

\vspace*{-4mm}
\section{The astronomical stability of the system}
The phase space of the \hr{} system appears strongly chaotic, with tiny islands
of regular two- and three-body MMRs. Hence, to study its long-term  (but finite)
evolution, we might rely on a notion of the {\em astronomical stability}
\citep{Lecar2001}, rather than on the formal Arnold's stability analyzed above. 
The astronomical stability may be investigated only by the direct numerical
integrations. Because the fate of chaotic configurations is hardly predictable, 
we attempted to gather statistics on initial conditions providing long-living
configurations, i.e., characterized by the {\em event time} $T_E$ of a close
encounter/ejection of a planet from an initial system. 
\begin{figure*}
\centerline{
        \hbox{\includegraphics[width=54mm]{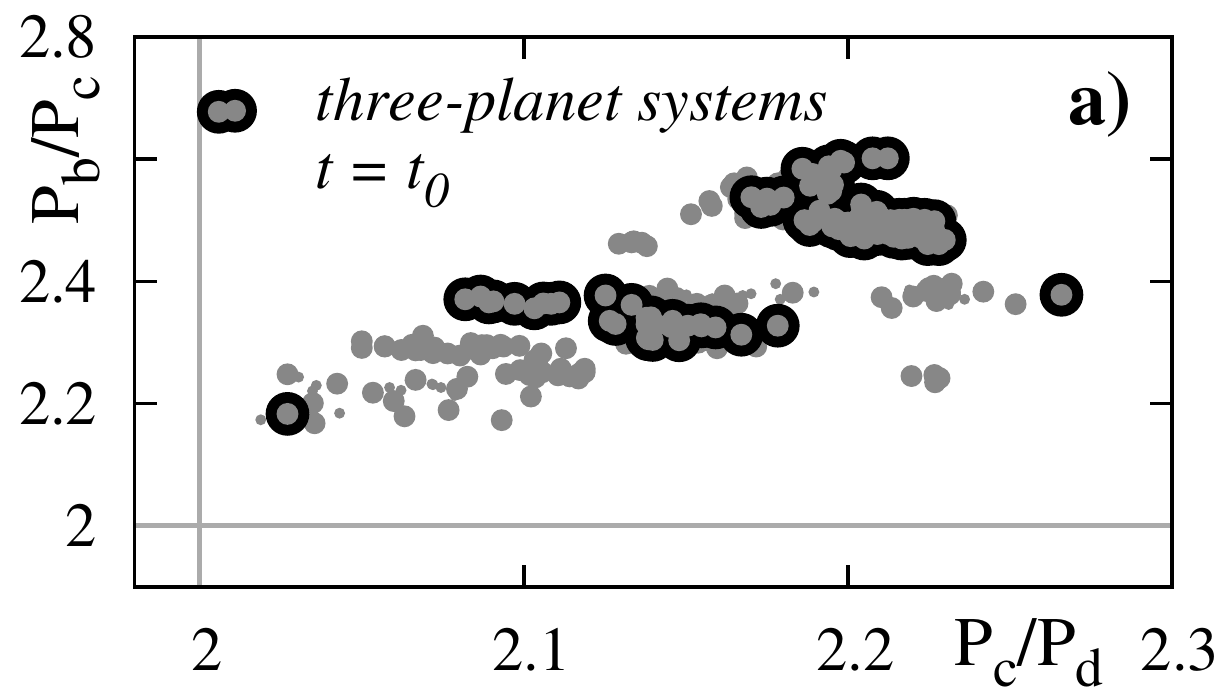} 
              \includegraphics[width=54mm]{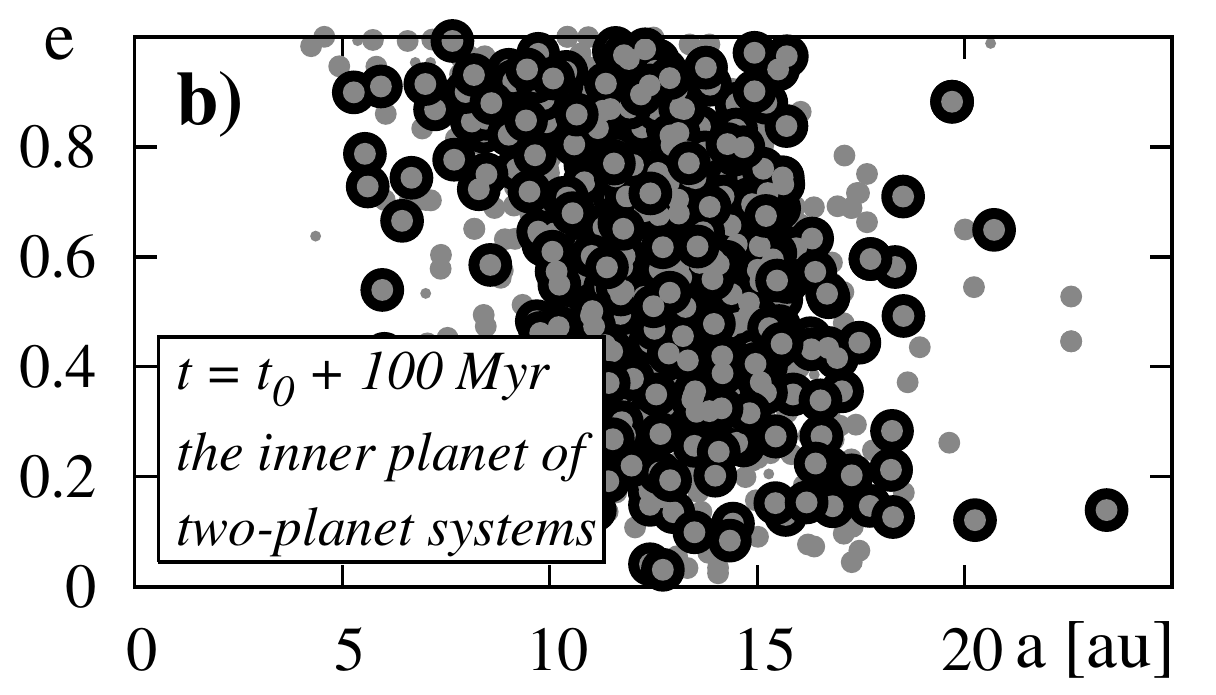} 
	      \includegraphics[width=54mm]{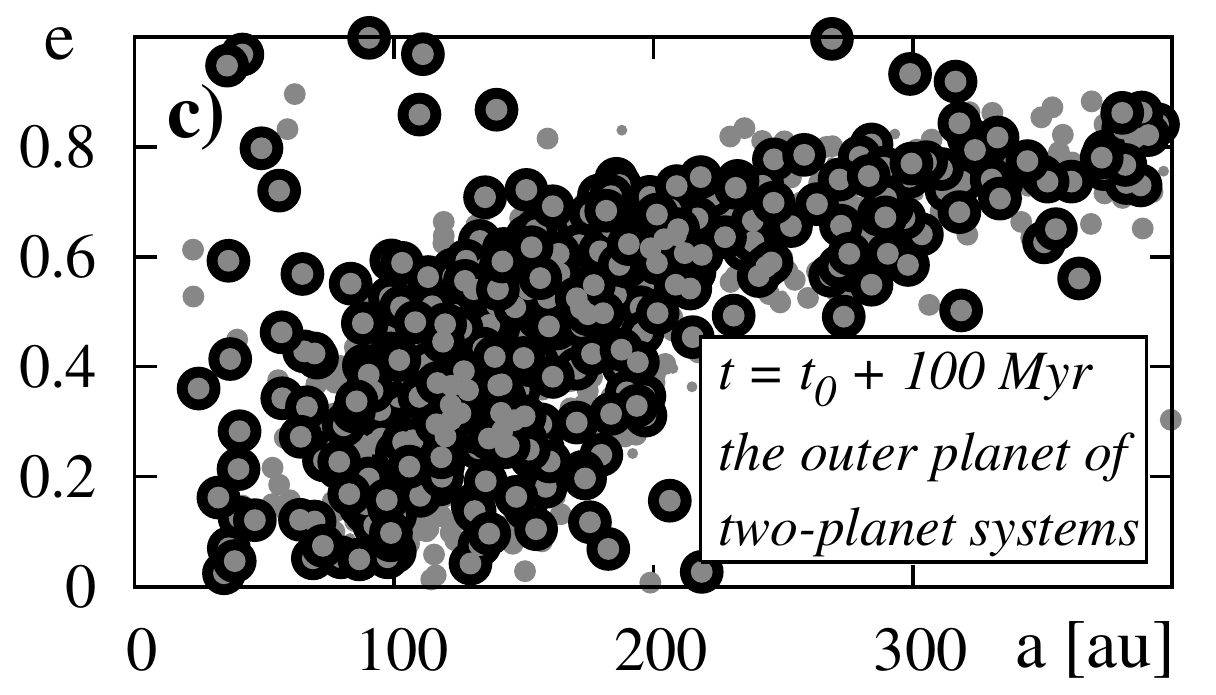}
	      }
}
\caption{
Osculating elements of configurations astronomically stable over $\Delta
t=+100$~Myr after the initial epoch $t_0$ in terms of orbital periods
ratio (panel a).  Panels (b) and (c) illustrate the distribution in the
$(a,e)$-plane of the final, dynamically relaxed systems of two planets.
}
\label{fig:fig7}
\end{figure*}

We tested initial conditions within formal $3\sigma$-level  of the nominal
Fit~I. To search for long-living systems, we again applied the GAMP algorithm,
with the penalty term $\Delta\Chi$  multiplied by $\tau=1-\|T_E/\Delta t\|$,
where $\Delta t$ is the maximal integration time relative to the initial
(present) epoch $t=t_0$.  In the first simulation, we integrated the system
over  $\Delta t = -30$~Myr (backwards), keeping track of solutions that survived
as three-planet configurations.  Next, the GAMP sample of $\sim 2000$ solutions
``living in the past'' was integrated up to $\Delta t = +100$~Myr. The results
are shown in Fig.~\ref{fig:fig7}. Panel~\ref{fig:fig7}a is for the orbital
periods ratio of systems that survived as three-planet configurations.  It
indicates that such systems are close to the 1d:2c MMR; the outer planets may be
also involved in 1c:2b~MMR or other low-order MMRs (like 2c:5b).  That agrees
well with  the results of \cite{Fabrycky2008}. Moreover, most of the tested
systems self-disrupted. Two remaining panels in Fig.~\ref{fig:fig7} are for the
final osculating elements in the two-planet sample. Due to intensive
planet--planet scattering (we recall that $K \sim 2$), the distribution spans
large ranges of semi-major axes and almost whole available range of
eccentricity. The strongly chaotic character of the \hr{} system leads to rapid
collisions/ejections in most of tested configurations during at most a few Myr.
That confirms globally that the dynamical maps shown in
Figs.~\ref{fig:fig3},\ref{fig:fig5} represent a generic picture of the phase
space, although they were computed for particular (resonant) initial conditions.
Still, only $\sim 400$ systems of the total population, i.e., less than 20\%,
survived the integrations. In fact, the sample is ``biased'' by the selection of
systems surviving the integrations backwards. We found that the direct Monte-Carlo
integrations leave much less than $1\%$ of astronomically stable configurations
after 100~Myr.  Hence, the self-adapting GAMP is crucial in this test because
the direct Monte-Carlo simulations would lead to unacceptable CPU overhead. 
Actually, there is no guarantee that systems astronomically stable in the
$100$~Myr test will also remain stable on longer time-scale, as shows the case
of Fit~III.  In  this experiment, we also found $\sim 20\%$ of single-planet
systems, and the rest in the sample ended as two-planet configurations. These
dynamically relaxed two-planet systems appear highly hierarchical, with a strong
maximum of semi-major axes ratio  $\alpha \sim 0.05$ (see also
Fig.~\ref{fig:fig7}b,c).

\section{Two-planet hypothesis}

{
Up to now, we assumed that the \hr{} hosts {\em three} planets.  Due to short
observations that revealed planet~d (a few weeks only), its orbit is
unconstrained. \cite{Marois2008} claim a $6\sigma$ detection of the common
proper motion, consistent with the Keplerian orbit. It is enforced by the
absence of \hr{}d in the HST images in 1998 \citep{Lafreniere2009} that
otherwise should be seen in the substracted light annulus (C. Marois, {\em
private comm.}). Still, we look here for an alternative explanation of the
strongly unstable \hr{} system due to projected brown dwarf or already ejected
planet that would be really too distant to influence orbits of \hr{}b,c.
}

We repeated the GAMP experiment for such two-planet model.  We found easily
rigorously stable solutions with $\Chi$ comparable with the nominal,  kinematic
best-fit system.  Elements of the best-fit solution are given in Table~1
(Fit~V).  The two-planet fits within $1\sigma$-bound span a wide range of
semi-major axes $\sim 10$~au and may be stable up to $e_{\idm{b,c}} \sim 0.15$.
Their orbits are initially close to anti-aligned ones. Planetary masses in such
systems remain in the 10~m$_{\idm{Jup}}$ range that is well consistent with
astrophysical constraints given in \citep{Marois2008}; we note that stable
three-body fits tend to much lower masses than declared \citep{Fabrycky2008}.
Also the dynamical map in Fig.~\ref{fig:fig8} shows extended zones of stability
and a proximity of the best-fit solution to the 5:2~MMR, recalling the
Jupiter-Saturn pair in the Solar-system. 

\begin{figure}
\centerline{
        \hbox{
	      \includegraphics[width=76mm]{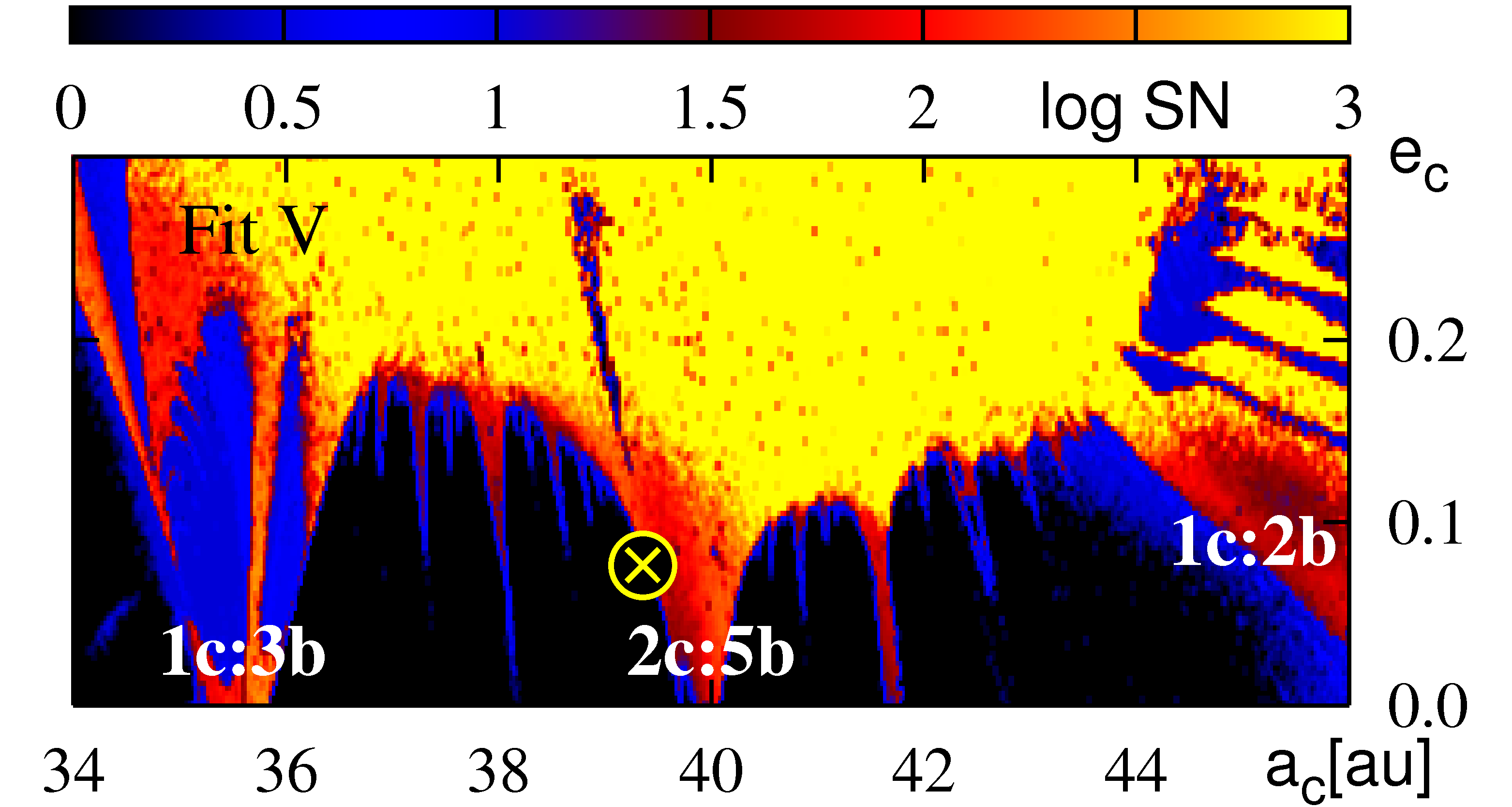}}
}
\caption{
Dynamical $(a_{\idm{c}},e_{\idm{c}})$-map of the best Fit~V (see Table~1) of two-planet
configuration in terms of the SN indicator. Its position is marked by a crossed
circle. Most prominent structures of low-order MMRs are labeled. 
}
\label{fig:fig8}
\end{figure}

\vspace*{-3mm}
\section{Conclusions}

The dynamical analysis of available astrometric data of  \hr{} reveal that its
massive companions are involved in heavy mutual interactions.   Assuming
$1\sigma$-range of the planetary and star mass astrophysical estimates, the
search for stable (regular) systems brings only narrow and very limited islands
of ordered motions. Most likely,  the system can be long-term stable if is
involved in low-order two- or three-body MMRs (particularly, in the Laplace
1d:2c:4b~MMR).  Here, we confirm the results of \cite{Fabrycky2008} and
\cite{Reidemeister2009}, which we derived after independent, quasi-global GAMP
calculations. Moreover, also peculiar 1d:1c~MMR Trojan systems stable over a
few~Gyr can be found.  

The outstanding discovery, in the light of  the dynamical analysis, brings a few
open questions. How the three-planet system may be captured in such tiny regions
of stable motions? Are in fact planetary masses much lower than estimated? Or is
the system substantially non-coplanar? Both these factors could extend the zones
of stability. While the masses may be constrained by astrophysical factors and
astrophysical-age estimates \citep{Marois2008,Reidemeister2009}, we can say
little  on the real mutual  inclinations. Further, if we ``skip'' the less
constrained object, the sub-system of outermost planets is stable,
resembling the Jupiter-Saturn pair, even if the masses are large, apparently
solving the puzzle. It may be verified soon, thanks to the shortest orbital
period of planet~d.

Actually, should we expect that the system is or must be stable? Its parent star
is very young, and we may have an opportunity to observe a system undergoing the
dynamical relaxation. The statistical analysis suggest, that the final fate  of
coplanar systems constrained by available astrometric data most likely will be
two-planet, highly hierarchical  configuration with eccentric orbits.  Our
calculations show that less than $20\%$ of systems stable in the past and
remaining in the neighborhood of the best stable Fit~III remain stable {\em
after} 100~Myr. Likely, even much less  number of configurations survive longer
time due to possible, chaotic effects of the three-body interactions [MMRs
overlapping, \citep{Murray2001}]. A conclusion of \cite{Fabrycky2008} may be
repeated here. Although the \hr{} has been directly imaged,  {the interpretation
of its images is very difficult and yet non-unique.  Longer observations are required to constrain
orbits of its planets.
}
\vspace*{-3mm}
\section{Acknowledgments}
{
We warmly thank Daniel Fabrycky for an informative review that greatly improved
the manuscript. Many thanks to Christian Marois for corrections and permission
to use images of HR 8799, and to Alexander Krivov and Ralph Neuh\"auser for a
discussion. This work is supported by the Polish Ministry of Science, through Grants
1P03D~021~29 and 92/N-ASTROSIM/2008/0.
}
\vspace*{-3mm}
\bibliographystyle{mn2e}
\bibliography{ms}
\end{document}